\documentclass[aps,prb,reprint]{revtex4-2}

\usepackage{graphicx}

\usepackage{bm}

\begin{document}

\title{Symmetry-dependent field-induced elastic renormalization in the quantum-critical medium-entropy alloy NiCoCr\(_{0.8}\)}

\author{Tadataka Watanabe$^1$}
\email{watanabe.tadataka@nihon-u.ac.jp}
\author{Mai Watanabe$^1$}
\author{Brian C. Sales$^2$}
\affiliation{$^1$Department of Physics, College of Science and Technology, Nihon University, Chiyoda, Tokyo 101-8308, Japan}
\affiliation{$^2$Materials Science and Technology Division, Oak Ridge National Laboratory, Oak Ridge, Tennessee 37831, USA}
\date{\today}

\begin{abstract}
We investigate the elastic response of cubic NiCoCr\(_{0.8}\), a quantum-critical medium-entropy alloy, by high-precision ultrasound velocity measurements. At zero magnetic field, the three symmetrically independent elastic moduli \(C_{11}\), \((C_{11}-C_{12})/2=C_t\), and \(C_{44}\) monotonically harden upon cooling without pronounced elastic softening or anomalies. In magnetic fields applied along [110], however, a minute but systematic elastic renormalization of the order of \(10^{-5}\) is resolved at low temperatures. The tetragonal \(E_g\) mode \(C_t\) softens with increasing field, whereas the derived bulk \(A_{1g}\) mode \(C_B\) hardens, revealing an opposite field response between the two strain-symmetry channels. Comparison with the previously reported quantum-critical–Fermi-liquid crossover scale shows that the field-induced elastic renormalization develops already in the quantum-critical regime and evolves continuously across the crossover region, without a sharp elastic anomaly or threshold behavior. These results demonstrate that symmetry-resolved ultrasound provides a sensitive probe of subtle strain-dependent changes in the electronic and magnetic free energy of a chemically disordered itinerant quantum-critical metal.
\end{abstract}

\maketitle

\section{Introduction}

Quantum criticality in metallic magnetic systems provides a platform for investigating the interplay among magnetic fluctuations, itinerant electrons, and disorder \cite{Lohneysen,Brando}. The medium-entropy alloy NiCoCr\(_x\) with 0.5 $\leq x \leq$ 1.2 is particularly notable in this context because it forms a chemically disordered face-centered-cubic (fcc) solid solution while maintaining chemical homogeneity from the centimeter to the nanometer scale \cite{Jin,Sales,Sales2}. An earlier study by Sales $et$ $al.$ revealed a quantum-critical (QC) region near \(x\simeq1\), characterized by anomalous low-temperature transport, magnetic, and thermodynamic properties \cite{Sales2}. Subsequent work demonstrated that NiCoCr\(_x\) with \(x\simeq0.8\) exhibits critical behavior consistent with a ferromagnetic QC point in the high-disorder limit \cite{Sales}. In addition, Cr introduces frustrated magnetic interactions on the chemically disordered fcc lattice \cite{Sales}. For NiCoCr\(_{0.8}\), application of a magnetic field ($H$) suppresses the QC magnetic fluctuations and drives the system toward a Fermi-liquid (FL) regime, with a characteristic crossover temperature \(T^*(H)\) \cite{Sales}. Sales $et$ $al.$ further pointed out that strain and pressure may provide additional tuning parameters for the QC behavior of NiCoCr\(_{0.8}\) \cite{Sales}. These observations raise the question of how the QC electronic and magnetic degrees of freedom couple to lattice strain and how this coupling evolves under magnetic field. More generally, theory has shown that when a critical soft mode couples linearly to strain, elastic degrees of freedom can play an essential role in quantum criticality \cite{Zacharias}.

Ultrasound velocity measurements provide a sensitive means of addressing such a question because an elastic modulus represents the response of the free energy to a strain of a particular symmetry \cite{Luthi}. Moreover, relative changes in sound velocity can be measured with a precision on the order of parts per million (ppm), allowing weak strain-coupled fluctuations and excitations to be detected \cite{Luthi}. Ultrasound studies have revealed pronounced elastic anomalies in metallic systems with itinerant or quantum critical behavior. Examples include the itinerant helimagnet MnSi, in which the elastic response evolves strongly with magnetic field near the magnetic transition \cite{Petrova}, the heavy-fermion metal CeRu\(_2\)Si\(_2\), where sound velocity and attenuation exhibit strong anomalies near the metamagnetic transition \cite{Bruls}, and the itinerant paramagnet UCoAl, which exhibits pronounced elastic softening near a magnetic critical endpoint \cite{Yoshizawa}. In the heavy-fermion compound YbPtSb, logarithmic elastic softening and its field evolution have been observed near a quantum critical point and associated with non-Fermi-liquid behavior \cite{Nakanishi}. Beyond the long-wavelength elastic response probed by ultrasound, inelastic neutron scattering has also revealed phonon softening and an anomalous phonon mode near the QC composition of Ca\(_{2-x}\)Sr\(_x\)RuO\(_4\) \cite{Moore}.

In our previous ultrasound studies of frustrated magnets, pronounced elastic anomalies have revealed a variety of magnetoelastic phenomena, including spin-Jahn-Teller fluctuations and spin-cluster excitations in MgCr\(_2\)O\(_4\) \cite{Watanabe}, coexisting orbital and spin-cluster fluctuations in MgV\(_2\)O\(_4\) \cite{Watanabe2}, and multiple lattice instabilities in the disordered pyrochlore magnet NaCaCo\(_2\)F\(_7\) \cite{Watanabe3}. Elastic softness associated with magnetostructural fluctuations has also been observed in the low-symmetry frustrated antiferromagnets \(A\)Ti\(_2\)O\(_5\) (\(A=\) Co, Fe) \cite{Watanabe4}. More recently, ultrasound measurements of the metallic frustrated antiferromagnet CrB\(_2\) revealed distinct elastic responses associated with transverse magnetoelastic fluctuations and longitudinal electron-lattice coupling \cite{Watanabe5}. These studies demonstrate that symmetry-resolved elastic measurements can distinguish different strain-coupled magnetic and electronic correlations.

NiCoCr\(_{0.8}\), however, differs qualitatively from the localized-moment frustrated magnets in which large elastic softening is commonly observed. Its magnetism is highly itinerant and develops in the presence of strong chemical disorder \cite{Sales}. Thus, a localized-moment description of elastic softening is not expected to provide an appropriate starting point. Nevertheless, the reported sensitivity of its magnetic susceptibility to strain \cite{Sales} suggests that the QC free energy retains a finite coupling to lattice degrees of freedom. It therefore remains an open question whether the field-induced evolution from the QC regime toward the FL regime produces a detectable elastic response and, in particular, whether such a response depends on strain symmetry.

In the present study, we investigate the elastic properties of single-crystalline NiCoCr\(_{0.8}\) by high-precision ultrasound velocity measurements. We determine the three symmetrically independent elastic moduli of the cubic crystal, \(C_{11}\), \((C_{11}-C_{12})/2=C_t\), and \(C_{44}\), and examine the low-temperature field dependence of \(C_{11}\) and \(C_t\) in magnetic fields up to 7 T applied along [110]. At zero magnetic field, all three elastic moduli monotonically harden upon cooling without pronounced elastic softening or anomalies. Under magnetic field, however, the tetragonal \(E_g\) mode \(C_t\) exhibits a systematic softening, whereas the derived bulk \(A_{1g}\) mode \(C_B\) hardens. Here, \(A_{1g}\) and \(E_g\) denote the corresponding strain-symmetry channels of the cubic crystal. The opposite field responses of these two strain-symmetry channels reveal a symmetry-dependent elastic renormalization. Comparison with the previously reported QC–FL crossover scale further shows that this elastic renormalization develops already in the QC regime and evolves continuously across the crossover region.

\section{Experimental}

Single crystals of NiCoCr$_{0.8}$ were grown by the optical floating-zone method as described in Ref. \cite{Sales}.
Ultrasound velocities were measured using the phase-comparison technique, which enables relative changes in sound velocity, and hence in elastic modulus, to be measured with a precision at the ppm level.
Longitudinal and transverse sound waves at a frequency of 30 MHz were generated and detected by LiNbO$_3$ transducers attached to parallel mirror surfaces of the single crystal.
We measured all three symmetrically independent elastic moduli of the cubic crystal: the compressive modulus $C_{11}$, the tetragonal shear modulus $(C_{11}-C_{12})/2=C_t$, and the trigonal shear modulus $C_{44}$.
From the measured $C_{11}$ and $C_t$, we also obtained the bulk modulus $C_B=(C_{11}+2C_{12})/3=C_{11}-(4/3)C_t$.
The $C_{11}$, $C_t$, and $C_{44}$ modes were measured using a longitudinal sound wave with propagation $\mathbf{k}\parallel[001]$ and polarization $\mathbf{u}\parallel[001]$, a transverse sound wave with $\mathbf{k}\parallel[110]$ and $\mathbf{u}\parallel[1\bar{1}0]$, and a transverse sound wave with $\mathbf{k}\parallel[110]$ and $\mathbf{u}\parallel[001]$, respectively.
The sound velocities of NiCoCr$_{0.8}$ measured at room temperature (300 K) were approximately 5630 m/s for $C_{11}$, 2570 m/s for $C_t$, and 4620 m/s for $C_{44}$.
Magnetic fields up to 7 T were applied along [110] for the field-dependent measurements presented below.

\begin{figure}[b]
\centering
\includegraphics[keepaspectratio,width=\linewidth]{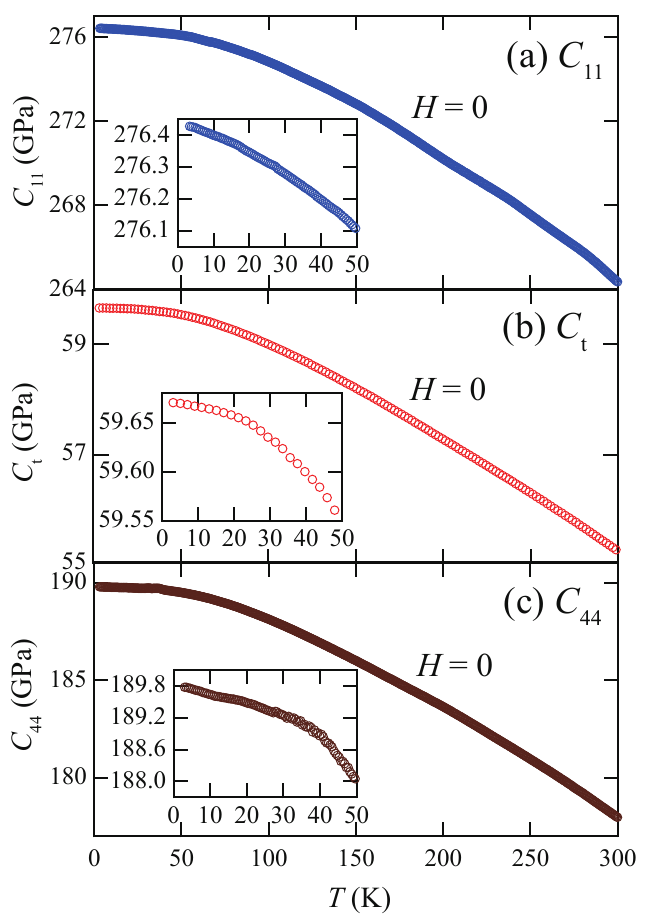}
\caption{Temperature dependence of the elastic moduli of NiCoCr$_{0.8}$ at zero magnetic field: (a) compressive modulus C$_{11}$, (b) tetragonal shear modulus $(C_{11}-C_{12})/2=C_t$, and (c) trigonal shear modulus $C_{44}$. The insets show expanded views below 50 K.}
\label{fig:fig1}
\end{figure}

\begin{figure}[t]
\centering
\includegraphics[keepaspectratio,width=\linewidth]{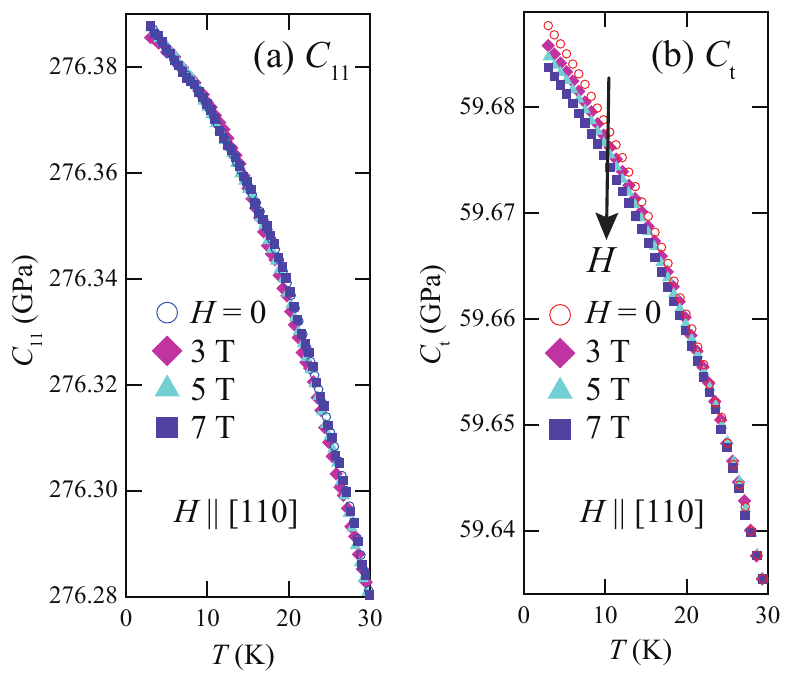}
\caption{Temperature dependence of (a) $C_{11}$ and (b) $C_t$ of NiCoCr$_{0.8}$ in magnetic fields of $\mu_0H=$ 0, 3, 5, and 7 T applied along [110]. The arrow in (b) indicates the variation of $C_t$ with increasing magnetic field.}
\label{fig:fig2}
\end{figure}

\section{Results}

Figure 1 depicts the temperature ($T$) dependence of the three symmetrically independent elastic moduli \(C_{11}(T)\), \(C_t(T)\), and \(C_{44}(T)\) of NiCoCr\(_{0.8}\) at zero magnetic field ($H$ = 0). All three elastic moduli monotonically harden upon cooling from 300 K to the lowest measured temperature of 3 K. As shown in the expanded views below 50 K [the insets in Fig. 1], no pronounced elastic softening or other anomalies are observed at low temperatures.

Figure 2 depicts the low-temperature behavior of \(C_{11}(T)\) and \(C_t(T)\) below 30 K in magnetic fields of \(\mu_0H=0\), 3, 5, and 7 T applied along [110]. The \(C_{11}(T)\) curves measured at different magnetic fields nearly overlap over the entire temperature range investigated [Fig. 2(a)]. In contrast, \(C_t(T)\) exhibits a small but systematic field-induced softening [Fig. 2(b)]. The magnitude of the softening increases with increasing magnetic field and becomes more pronounced toward low temperatures.

To examine these small field-induced changes more closely, Fig. 3 shows the relative changes \(\Delta C_\Gamma/C_\Gamma=[C_\Gamma(T,H)-C_\Gamma(T,0)]/C_\Gamma(T,0)\) for \(C_{11}\) and \(C_t\). For \(C_{11}\), the relative changes remain of the order of 10 ppm or less, and no systematic monotonic field dependence comparable to that of \(C_t\) is resolved [Fig. 3(a)]. In contrast, \(\Delta C_t/C_t\) is negative and systematically decreases with increasing magnetic field [Fig. 3(b)]. At \(T=3\) K, \(\Delta C_t/C_t\) reaches approximately \(-30\), \(-50\), and \(-65\) ppm for \(\mu_0H\) = 3, 5, and 7 T, respectively. The magnitude of the field-induced softening decreases continuously with increasing temperature and becomes small near the upper end of the temperature range investigated.

The weak field dependence of \(C_{11}(T)\) [Fig. 2(a)], together with the systematic field-induced softening of \(C_t(T)\) [Fig. 2(b)], results in an opposite field response of the bulk modulus \(C_B(T)\). Figure 4(a) shows \(C_B(T)=C_{11}(T)-(4/3)C_t(T)\). To evaluate \(C_B(T)\) at the same temperature points as the measured \(C_t(T)\) data, smooth fitting curves to the measured \(C_{11}(T)\) data at each magnetic field were used for interpolation. In contrast to \(C_t(T)\), \(C_B(T)\) hardens with increasing magnetic field. This behavior is more clearly seen in \(\Delta C_B/C_B\) shown in Fig. 4(b), which is positive and increases toward low temperatures and with increasing field. At $T$ = 3 K, the relative hardening is approximately 13, 20, and 27 ppm for \(\mu_0H\) = 3, 5, and 7 T, respectively. Figure 4(c) summarizes the field dependence of \(\Delta C_B/C_B\) and \(\Delta C_t/C_t\) at $T$ = 3 K. The bulk \(A_{1g}\) mode \(C_B\) and tetragonal \(E_g\) mode \(C_t\) exhibit field-induced elastic changes of opposite sign, and the magnitudes of both responses increase continuously with magnetic field.

\begin{figure}[t]
\centering
\includegraphics[keepaspectratio,width=\linewidth]{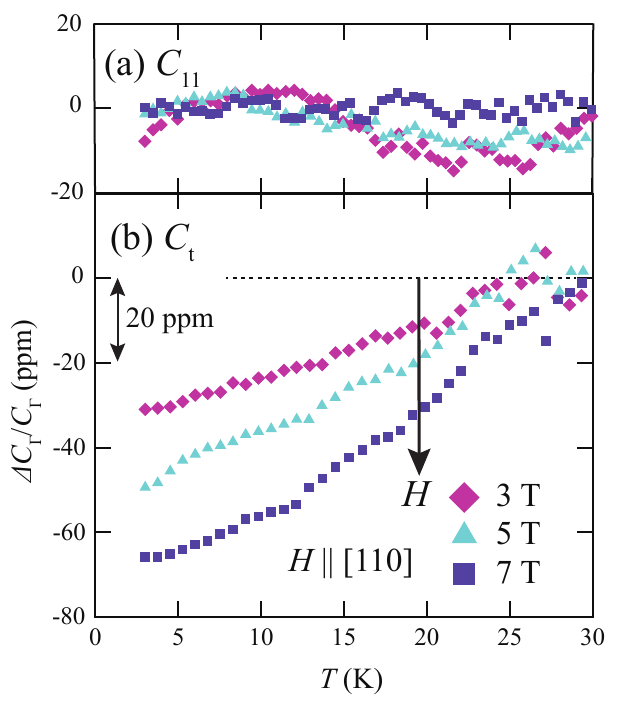}
\caption{Temperature dependence of the relative field-induced changes in (a) $C_{11}$ and (b) $C_t$ of NiCoCr$_{0.8}$ for $\mu_0H$ = 3, 5, and 7 T applied along [110]. The relative change is defined as $\Delta C_\Gamma/C_\Gamma=[C_\Gamma(T,H)-C_\Gamma(T,0)]/C_\Gamma(T,0)$, where $\Gamma=11$ or $t$. The arrow in (b) indicates the variation of $\Delta C_t/C_t$ with increasing magnetic field, and the double arrow denotes a scale of 20 ppm. The horizontal dotted line in (b) indicates $\Delta C_t/C_t=0$.}
\label{fig:fig3}
\end{figure}

\begin{figure*}[t]
\centering
\includegraphics[keepaspectratio,width=\linewidth]{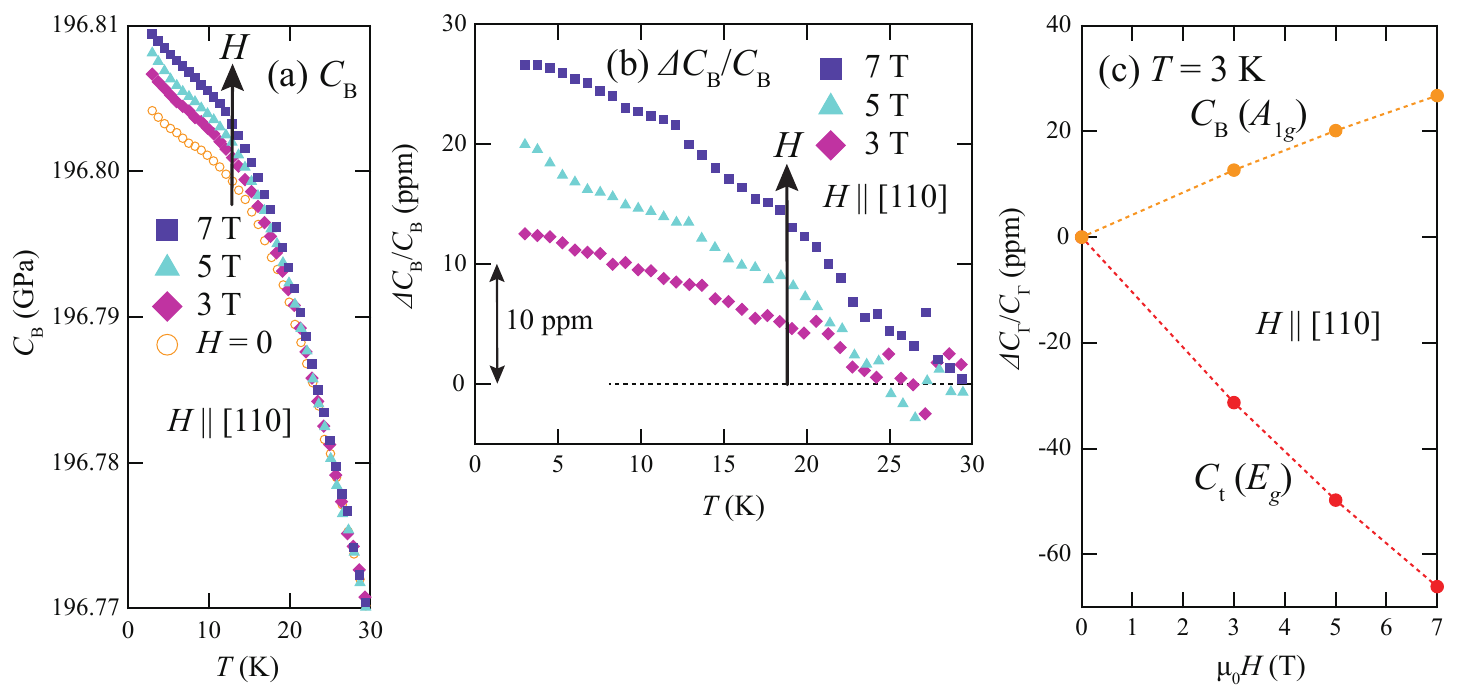}
\caption{Field-induced response of the bulk and tetragonal elastic modes of NiCoCr$_{0.8}$ for $H\parallel[110]$. (a) Temperature dependence of the bulk modulus $C_B$, calculated from the measured $C_{11}$ and $C_t$ using $C_B=(C_{11}+2C_{12})/3=C_{11}-(4/3)C_t$ at $\mu_0H$ = 0, 3, 5, and 7 T. The arrow indicates the variation of $C_B$ with increasing magnetic field. (b) Relative field-induced change $\Delta C_B/C_B=[C_B(T,H)-C_B(T,0)]/C_B(T,0)$. The arrow indicates the variation of $\Delta C_B/C_B$ with increasing magnetic field, and the double arrow denotes a scale of 10 ppm. The horizontal dotted line indicates $\Delta C_B/C_B=0$. (c) Magnetic-field dependence of the relative elastic change $\Delta C_\Gamma/C_\Gamma$ at $T$ = 3 K for the bulk $A_{1g}$ mode $C_B$ and the tetragonal $E_g$ mode $C_t$. The dotted lines are guides to the eye.}
\label{fig:fig4}
\end{figure*}

\section{Discussion}

The present ultrasound measurements reveal a minute but systematic field-induced elastic renormalization in NiCoCr\(_{0.8}\). An elastic modulus associated with a strain \(\varepsilon_\Gamma\) can generally be expressed as the curvature of the free energy with respect to that strain, \(C_\Gamma=\partial^2F/\partial\varepsilon_\Gamma^2\). The observed field dependence of the elastic moduli therefore demonstrates that the field-dependent electronic and magnetic free energy contributes to the elastic response through its strain dependence. This strain sensitivity is noteworthy in view of the previous observation that severe cold working strongly enhances the low-temperature magnetic susceptibility of NiCoCr\(_{0.8}\), which led Sales $et$ $al.$ to suggest strain and pressure as possible tuning parameters of its QC behavior \cite{Sales}. Although the present ultrasound measurements do not constitute a static strain-tuning experiment, they directly probe the strain dependence of the field-dependent free energy through the elastic response. In contrast to localized-moment magnetic systems, in which pronounced softening is often observed, NiCoCr$_{0.8}$ shows no such softening in its zero-field elastic moduli, and the field-induced changes are only of the order of 10 ppm. The small magnitude of the response is consistent with, although by itself does not establish, the highly itinerant character of the magnetism previously reported for NiCoCr\(_{0.8}\) \cite{Sales}. Accordingly, the present field-induced elastic response should not be interpreted in terms of a localized-moment elastic instability.

More importantly, the field-induced elastic renormalization is strongly dependent on strain symmetry. The bulk modulus \(C_B\), corresponding to the isotropic \(A_{1g}\) strain channel, hardens with magnetic field, whereas the tetragonal shear modulus \(C_t\), corresponding to the \(E_g\) channel, softens. Since \(C_{11}=C_B+(4/3)C_t\), the weak field dependence of \(C_{11}\) reflects a partial cancellation between the positive field-induced change in \(C_B\) and the negative field-induced change in \(C_t\). We emphasize that \(C_B\) is derived from the measured \(C_{11}\) and \(C_t\) and therefore does not constitute an additional independent measurement. Nevertheless, the decomposition of the directly measured elastic response demonstrates that the field-induced renormalization is not uniform among the different strain-symmetry channels.

This symmetry-dependent elastic response can be compared with the field-induced crossover from the QC regime to the FL regime reported in Ref. \cite{Sales}. Sales $et$ $al.$ showed that application of a magnetic field suppresses magnetic fluctuations and tunes NiCoCr\(_{0.8}\) toward a normal FL, with the crossover temperature \(T^*\) determined from the maximum in the temperature dependence of the magnetic susceptibility. Expressed in the present field notation, the reported relation \(T^*=5.62(\mu_0H)^{0.67}\), with \(\mu_0H\) in tesla, gives \(T^*\simeq11.7\), 16.5, and 20.7 K for \(\mu_0H=3\), 5, and 7 T, respectively. At these crossover temperatures, the present elastic data give \(\Delta C_t/C_t\simeq-21.4\), \(-24.2\), and \(-28.3\) ppm, respectively, while the corresponding values of the derived bulk response are \(\Delta C_B/C_B\simeq8.7\), 9.8, and 11.4 ppm.

Interestingly, the magnitude of the bulk elastic renormalization at \(T^*(H)\) remains of the same order, \(\Delta C_B/C_B\sim10\) ppm, for all three magnetic fields. This empirical correspondence may suggest that the QC–FL crossover occurs at a comparable degree of renormalization in the isotropic \(A_{1g}\) strain channel. We emphasize, however, that \(\Delta C_B/C_B\) represents a change in the curvature of the free energy with respect to strain and therefore cannot be directly converted into an equivalent static strain or pressure. Direct hydrostatic-pressure or static-strain measurements of \(T^*\) would be valuable for clarifying this correspondence. 

More generally, the field-induced elastic renormalization is already finite on approaching \(T^*(H)\) from the QC side, and no sharp anomaly is resolved in either \(C_t(T)\) or \(C_B(T)\) at \(T^*(H)\). Instead, the field-induced changes develop continuously over a broad temperature range that spans the QC–FL crossover. Such behavior is consistent with the crossover character of \(T^*(H)\), rather than a thermodynamic phase boundary. The present results therefore do not indicate an elastic threshold associated with the onset of the FL regime. Rather, they show that the QC–FL crossover occurs within a $T-H$ region in which a finite magnetoelastic renormalization has already developed. This correspondence suggests a close connection between the evolution of the strain-sensitive magnetic/electronic correlations and the QC–FL crossover, although the present measurements do not establish a causal relationship between them.

The opposite responses of the \(A_{1g}\) and \(E_g\) elastic channels further indicate that the field-induced evolution cannot be described simply as a uniform suppression of all strain-coupled magnetic correlations. NiCoCr\(_{0.8}\) is a chemically disordered itinerant magnet on an fcc lattice, in which Cr introduces frustrated magnetic interactions \cite{Sales}. First-principles calculations for equiatomic CrCoNi have further shown that the distribution of Cr magnetic moments depends strongly on the local atomic environment and differs between fcc and hexagonal close-packed (hcp) configurations, indicating a sensitivity of the magnetic state to the local lattice environment \cite{Niu}. Although these calculations concern structural phase stability and large shear deformations rather than the small-strain elastic response probed here, they illustrate the sensitivity of the magnetic state to structural and local chemical environments in the Ni–Co–Cr alloy system.

For NiCoCr\(_{0.8}\), one possible interpretation is that magnetic field reorganizes the strain-coupled magnetic/electronic correlations in a symmetry-dependent manner, possibly through changes in the relative contributions from different local or bond environments. Such a redistribution could naturally produce different elastic renormalizations in the isotropic and tetragonal strain channels.  Although the microscopic origin of this symmetry-dependent response cannot be determined from the present ultrasound measurements alone, the symmetry-resolved response itself provides information beyond that available from bulk magnetic measurements and demonstrates the sensitivity of high-precision ultrasound to minute magnetoelastic effects in a chemically disordered itinerant QC metal.

The present results may also have broader implications for medium- and high-entropy magnetic materials, in which strong chemical disorder produces a wide distribution of local atomic and magnetic environments while the average crystallographic symmetry can remain high. Symmetry-resolved ultrasound measurements may provide a useful means of detecting how such local magnetic correlations are reorganized by external fields even when the associated elastic response is extremely small.

\section{Summary}

In summary, we investigated the elastic properties of the quantum-critical medium-entropy alloy NiCoCr\(_{0.8}\) by high-precision ultrasound velocity measurements. At zero magnetic field, all three symmetrically independent elastic moduli \(C_{11}\), \(C_t\), and \(C_{44}\) monotonically harden upon cooling without showing pronounced elastic softening or anomalies. In magnetic fields applied along [110], however, a minute but systematic elastic renormalization was resolved at low temperatures. The tetragonal \(E_g\) mode \(C_t\) softens with increasing magnetic field, whereas the derived bulk \(A_{1g}\) mode \(C_B\) hardens, demonstrating a symmetry-dependent magnetoelastic response. Comparison with the quantum-critical–Fermi-liquid crossover scale reported previously shows that this field-induced elastic renormalization is already finite in the quantum-critical regime and develops continuously across the crossover region, without a sharp elastic anomaly or threshold behavior. These results demonstrate that symmetry-resolved ultrasound can detect subtle strain-dependent changes in the electronic and magnetic free energy of a chemically disordered itinerant quantum-critical metal and provide a sensitive probe of magnetoelastic correlations in medium- and high-entropy magnetic materials.

\begin{acknowledgments}
This work was partly supported by JSPS KAKENHI Grant No. JP21K03476.
\end{acknowledgments}

\end{document}